\documentclass[11pt,reqno]{amsart}

\usepackage[T1]{fontenc}
\usepackage{lmodern}
\usepackage{microtype}
\usepackage[margin=1.15in]{geometry}

\usepackage{mathtools}
\usepackage{amssymb,amsfonts,amsthm}
\usepackage{mathrsfs}

\usepackage{graphicx}
\usepackage{subcaption}
\usepackage{booktabs}
\usepackage{float}

\usepackage[numbers,square,sort&compress]{natbib}
\usepackage[colorlinks=true,linkcolor=blue,citecolor=blue,urlcolor=blue]{hyperref}
\usepackage[capitalize,noabbrev]{cleveref}

\numberwithin{equation}{section}

\theoremstyle{plain}
\newtheorem{theorem}{Theorem}[section]

\theoremstyle{definition}

\theoremstyle{remark}

\title{Phylogenetic Tree Inference with Tropical Axial Attention}

\author{Chris Teska}
\author{Kurt Pasque}
\author{Ruriko Yoshida}

\author{Baran Hashemi}

\begin{document}

\begin{abstract}
 In this work, we introduce a Tropical Axial Attention neural reasoning architecture that replaces vanilla softmax dot-product attention with max-plus operators, inducing a piecewise-linear structure aligned with dynamic programming formulations. From multi-species sequence alignments, our model learns all possible pairwise distances and is trained using a combination of $\ell_1$ and tropical symmetric distance metric losses with an ultrametric violation penalty. We leverage the well known isomorphic relationship between the space of all phylogenetic trees with $n$ species and tropical Grassmannian to show that tropical attention provides a natural geometric framework for phylogenetic inference.
  On empirical $DS1-DS11$ alignments, where true trees are unknown, the tropical model {achieves the lowest MAE to its FastME-induced tree metric on every dataset, with a MAE reductions averaging 81.5\% relative to Phyloformer and 98.4\% relative than Phyloformer 2}. These results suggest that tropical attention is a useful geometric inductive bias for neural phylogenetic inference, especially under distribution shift and when tree-metric consistency is important.

 \end{abstract}

\maketitle

\section{Introduction}
A phylogenetic tree represents the evolutionary history among a set of species, genes, or sequences, and it is used in a wide range of applications, from the comparative analysis of
genes \citep{Soltis2003} to the analysis of Coronavirus disease \citep{covid}. Phylogenetic tree reconstruction is a combinatorial optimization problem: given a multiple sequence alignment (MSA), the goal is to infer a tree that explains the observed evolutionary relationships. Standard statistical approaches include Maximum Likelihood (ML) \citep{felsenstein1981evolutionary} and Bayesian Inference \citep{MrBayes}. Distance-based methods form another widely used class of approaches. For example, Neighbor Joining (NJ) \citep{saitou1987neighbor} and Balanced Minimum Evolution (BME) \citep{BME,lefort2015fastme} {takes an estimated} pairwise distance matrix from an MSA {(for example, via the stochastic speciation process \citep{kuhner1994simulation})} and then infer a tree from that matrix.
The second step in a distance-based method is difficult because the space of unrooted phylogenetic trees is a union of $(n-3)$-dimensional polyhedral cones in $\mathbb{R}^{\binom{n}{2}}$ \citep{Billera2001}. In fact, estimating a phylogenetic tree via BME is NP-hard \citep{day1987computational} {and also APX-hard \citep{FIORINI201231}. It is also} well-known that NJ is an approximation to BME \citep{NJ-BME}.  {For more details see \citep{BME2}.  One can find a tutorial on geometric approaches to phylogenetic tree reconstruction in \citep{HAWS2026343}.  }

{A key for a successful phylogenetic tree estimation from an input MSA via a distance-based method is to estimate a good distance matrix from the MSA.  }
Recent neural approaches {to estimate a distance matrix from a MSA}, including Phyloformer \citep{nesterenko2025phyloformer} and Phyloformer 2 (EvoPF) \citep{blassel2025likelihood}, replace explicit search with transformer models that take MSAs as input and output distance matrices or tree distributions. These models operate in Euclidean representations and rely on standard attention mechanisms, which may not fully capture the combinatorial structure inherent in phylogenetic inference \citep{vaswani2023attentionneed}. {For interested readers, definitions and basic concepts on LLM reasoning models can be found at \citep{wang2025tutorialllmreasoningrelevant}.}

In this work, {we focus on the problem to estimate a distance matrix of all pairwise distances from an input MSA to as close to a tree metric as possible. Here,} we propose a geometrically motivated alternative: we replace softmax dot-product attention with tropical (max-plus) attention \citep{hashemi2025tropical}, yielding a piecewise-linear architecture that is better aligned with the dynamic-programming and combinatorial-optimization structure of tree reconstruction. Tropical algebra provides a natural framework for representing the polyhedral constraints that arise in distance-based phylogenetic inference.

We introduce a Tropical Axial Transformer, which:
\begin{itemize}
    \item applies tropical attention over pairwise sequence representations,
    \item uses axial decomposition over sequence and pair dimensions,
    \item and enforces structure through losses grounded in tropical geometry.
\end{itemize}

{These structural choices are reflected empirically in competitive overall performance and improved branch-length sensitive reconstruction and tree-metric consistency under distribution shift relative to Phyloformer baselines.}

\section{Background}
\subsection{Tropical Algebra Primer}

Tropical algebra provides an alternative arithmetic in which addition and multiplication are replaced by max and addition, respectively. This algebraic structure is appropriate for problems involving piecewise-linear geometry, combinatorial optimization, and max constraints, all of which arise naturally in phylogenetics.

The \emph{max-plus tropical semiring} is defined as the set $\mathbb{R} \cup \{-\infty\}$ equipped with the operations
\[
a \oplus b = \max(a,b), \qquad a \odot b = a + b,
\]
for $a, \, b, \, c \in \mathbb{R}\cup \{-\infty\}$.
In this setting, tropical addition corresponds to taking a maximum, while tropical multiplication corresponds to ordinary addition. These operations are associative, commutative, and satisfy distributivity:
\[
a \odot (b \oplus c) = (a \odot b) \oplus (a \odot c)
\]
for $a, \, b, \, c \in \mathbb{R}\cup \{-\infty\}$.
This algebra produces functions that are piecewise linear rather than smooth. Since it operates over polyhedral structures vice points in space, tropical geometry naturally describes combinatorial geometric structures over \emph{sets of inputs} which generate polyhedral structure, such as shortest paths, dynamic programming recursions, and tree reconstruction \citep{hashemi2025tropical, joswig2022parametric, jukna2014lower, axiotis2018capacitated}.

\subsection{Tree Space}
Phylogenetic trees are the fundamental mathematical representation of evolutionary relationships between species or taxa, yet the space of all possible trees is both combinatorially vast and geometrically complex. Formally, a phylogenetic tree $G \in \mathcal{T}_n$ is a connected, acyclic graph with $n$ leaves with labels, internal nodes represent ancestral divergence without labels, and edges assigned nonnegative weights corresponding to evolutionary distances. In the rooted binary case, each internal node has degree three (one parent and two children), and the number of distinct rooted phylogenetic tree topologies on $n$ labeled leaves is given by
\[
(2n - 3)!! = (2n - 3)(2n - 5)\cdots 3 \cdot 1,
\]
which grows super-exponentially with $n$. Even for relatively small $n$, this space becomes astronomically large, making exhaustive search computationally infeasible and making inference a fundamentally difficult problem \citep{bhatt2023tropical}. This combinatorial explosion is compounded by the fact that phylogenetic trees are not naturally embedded in a standard Euclidean space, but instead form a structured, non-Euclidean subspace inside Euclidean space.

A major step toward understanding this structure was the introduction the Billera-Holmes-Vogtmann (BHV) construction \citep{Billera2001}. In this framework, each tree is represented by its internal edge lengths, with each topology corresponding to a Euclidean region (orthant) of dimension $n-3$, reflecting the number of internal edges in an unrooted binary tree \citep{ren2021combinatorial}. These regions are glued together along lower-dimensional faces corresponding to degenerate trees obtained by collapsing edges (internal edge length $0$). While this provides a rigorous geometric foundation with desirable properties, the resulting space is only locally Euclidean. In particular, the presence of non-smooth boundaries complicates optimization and statistical analysis, and averages of trees may lie on these boundaries, producing unresolved topologies.  Although BHV space provides an elegant description, it does not fully resolve the challenges posed by the geometry of phylogenetic trees \citep{monod2022tropical}.

An alternative perspective arises from tropical geometry, which provides a fundamentally different way to represent and analyze tree space. In particular, \cite{speyer2004tropical} showed that the space of phylogenetic trees with $n$ leaves is isomorphic to the tropical Grassmannian. Tropical geometry replaces classical addition and multiplication with max-plus (or min-plus) operations, transforming algebraic objects into piecewise-linear structures \citep{maclagan2015tropical}. Under this framework, polynomials become maxima or minima of linear functions, and geometric objects become polyhedral complexes. This shift from smooth to piecewise-linear geometry aligns naturally with the structure of the space of phylogenetic trees, which is governed by combinatorial constraints and extremal relationships. In particular, tree metrics are characterized by inequalities such as the {\em four-point condition}, which depend on maximum operations rather than linear combinations.

This connection becomes precise through the tropical Grassmannian, which provides a complete characterization of phylogenetic tree space in terms of tropical algebra. In this formulation, tree metrics correspond to points in a tropical variety defined by tropical polynomials, linking phylogenetics directly to algebraic geometry. Unlike Euclidean representations, which struggle to capture the discrete and hierarchical nature of trees, the tropical framework encodes these properties directly through its operations and polyhedral structure. Moreover, methods with tropical operations are closely connected to optimization and dynamic programming, where similar max-plus operations arise naturally, offering both computational and theoretical advantages \citep{maclagan2015tropical}.

Taken together, these observations suggest that the difficulty of phylogenetic inference is not merely computational, but structural. Tree space is inherently non-Euclidean, combinatorial, and governed by extremal relationships. Tropical geometry provides a natural mathematical language for this setting, offering representations that align with both the geometry and combinatorics of phylogenetic trees. This perspective motivates the use of tropical methods, including tropical neural architectures, as a principled approach to modeling and inference in phylogenetics.

\subsection{Tree Inference}

A standard approach to phylogenetic inference is to represent a tree through pairwise distances between its leaves. Let $G \in \mathcal{T}_n$ be a phylogenetic tree with $n$ leaves and nonnegative edge weights. For each pair of leaves $(i,j)$, define
\[
d_{ij} = \text{length of the unique path between } i \text{ and } j.
\]
Collecting these values yields a symmetric, zero diagonal matrix $d = (d_{ij}) \in \mathbb{R}^{\binom{n}{2}}$,
whose coordinates correspond to unordered leaf pairs $(i,j)$.

This representation embeds phylogenetic trees into an ambient Euclidean space of pairwise distances. However, not every vector $d \in \mathbb{R}^{\binom{n}{2}}$ corresponds to a valid tree. The subset that does is known as the space of \emph{tree metrics}, which is defined by additional combinatorial constraints arising from the branching structure of trees \citep{monod2022tropical}.

The fundamental characterization of tree metrics is given by the four-point condition.

\begin{theorem}[Four-Point Condition]\label{thm:fourpoint}
A vector $d \in \mathbb{R}^{\binom{n}{2}}$ represents a phylogenetic tree metric if and only if, for all distinct indices $i,j,k,\ell$,
\[
d_{ij} + d_{k\ell} \leq \max\bigl(d_{ik} + d_{j\ell}, \; d_{i\ell} + d_{jk}\bigr),
\]
and the maximum among the three quantities
\[
d_{ij} + d_{k\ell}, \quad d_{ik} + d_{j\ell}, \quad d_{i\ell} + d_{jk}
\]
is attained at least twice.
\end{theorem}

Proofs for this theorem are widely available in phylogenetic textbooks. \cite{speyer2004tropical} showed this condition is equivalent to the tropical Pl\"ucker relations defining the tropical Grassmannian $\mathrm{trop}(\mathrm{Gr}(2,n))$. In particular, the expressions
\[
\max\left\{d_{ij} + d_{k\ell}, \quad d_{ik} + d_{j\ell}, \quad d_{i\ell} + d_{jk}\right\}
\]
are tropical linear forms, and the requirement that the maximum is attained at least twice corresponds to membership in a tropical variety. For more detail, see \cite{maclagan2015tropical} and \cite{monod2022tropical}. Thus, the space of phylogenetic tree metrics is naturally identified with a subset of the tropical Grassmannian, linking tree inference directly to tropical algebraic geometry.

A particularly important subclass of tree metrics is given by ultrametrics, corresponding to rooted trees with equal root-to-leaf distances.

\begin{theorem}[Ultrametric Condition]\label{thm:ultrametric}
A vector $d \in \mathbb{R}^{\binom{n}{2}}$ is an ultrametric ($d \in \mathcal{U}_n$) if and only if, for all distinct $i,j,k$,
\[
d_{ij} \leq \max\bigl(d_{ik}, d_{jk}\bigr),
\]
and the maximum among $\{d_{ij}, d_{ik}, d_{jk}\}$ is attained at least twice.
\end{theorem}

The ultrametric condition is both necessary and sufficient for the existence of a rooted equidistant tree and implies $\mathcal{U}_n \subset \mathcal{T}_n$. Geometrically, this enforces a hierarchical structure in which distances are determined by the height of the least common ancestor.

\subsubsection{Tropical Algebra and Tree Inference}

A central feature of tropical geometry is invariance under additive shifts. For a vector $x \in \mathbb{R}^D$, adding a constant $c$ to all coordinates does not change its essential structure. This leads to the tropical projective torus \citep{monod2022tropical}
$\mathbb{R}^D / \mathbb{R}\mathbf{1}$, where $\mathbf{1} = (1,\dots,1)$ and, in the phylogenetic setting, $D = \binom{n}{2}$ indexes all pairwise distances between $n$ taxa.

Two vectors $x,y \in \mathbb{R}^D/ \mathbb{R}\mathbf{1}$ are equivalent if
\[
x \sim y \quad \Longleftrightarrow \quad x = y + c \cdot \mathbf{1}
\]
for some $c \in \mathbb{R}$, that is to say whenever their coordinate-wise differences are constant.

In the context of phylogenetics, pairwise distance representations of trees naturally lie in this quotient space. Adding a constant to all pairwise distances corresponds to a uniform shift in evolutionary time and does not alter the underlying tree topology. Consequently, both tree metrics and ultrametrics are more precisely viewed as elements of $\mathbb{R}^D/ \mathbb{R}\mathbf{1}$ rather than $\mathbb{R}^D$.

A natural notion of distance on the quotient space is given by the \emph{Hilbert projective metric} \citep{maclagan2015tropical, hashemi2025tropical}, also referred to as the tropical metric. For $x,y \in \mathbb{R}^D / \mathbb{R}\mathbf{1}$, it is defined as
\[
d_{tr}(x,y)
= \max_i (x_i - y_i) - \min_i (x_i - y_i).
\]

This metric has several important properties \citep{hashemi2025tropical}:
\begin{itemize}
\item \textbf{Projective invariance:} For any $c \in \mathbb{R}$,
\[
d_{tr}(x + c\mathbf{1}, y+ c\mathbf{1}) = d_{tr}(x,y),
\]
so it is well-defined on $\mathbb{R}^{D}/\mathbb{R}\mathbf{1}$.

\item \textbf{Symmetry:} $d_{tr}(x,y) = d_{tr}(y,x)$.

\item \textbf{Non-negativity:} $d_{tr}(x,y) \geq 0$, with equality if and only if $x \sim y$.

\item \textbf{Value invariance:} The metric depends only on the spread of coordinate-wise differences, rather than their absolute values.
\end{itemize}

In phylogenetic inference pairwise distances are often treated as elements of a Euclidean space, but the space of valid tree metrics is not closed under linear operations and is instead defined by the polyhedral structure of tree space. This mismatch introduces distortions when Euclidean operations and loss functions are used for learning.

In contrast, tropical geometry provides a representation in which this structure is naturally expressed. Tree metrics form a structured subset of $\mathbb{R}^{\binom{n}{2}}/\mathbb{R}\mathbf{1}$, and their defining relations—such as the four-point condition—are governed by tropical algebra. Within this space, the Hilbert projective metric provides a geometry that is invariant to data shifts and sensitive to relative structure making tropical algebra ideal for phylogenetic inference.

\section{Related Work}
Practitioners have used specialized statistical and combinatorial methods for much of the history of phylogenetic tree inference. ML approaches estimate the tree and branch lengths that maximize the probability of the observed sequence alignment under a specified substitution model \citep{felsenstein1981evolutionary}. These methods operate directly on sequence data but require searching a vast, discrete space of tree topologies, making them computationally expensive despite heuristic optimizations used in tools such as RAxML and IQ-TREE \citep{stamatakis2014raxml, nguyen2015iqtree}. 

Distance-based methods instead first {takes an estimated pairwise distance matrix from the MSA and} then construct a tree consistent with these distances \citep{saitou1987neighbor,lefort2015fastme, price2010fasttree}. Geometrically, distance-based methods can be interpreted as projecting an estimated pairwise distance matrix from $\mathbb{R}^{\binom{n}{2}}$ onto the space of tree metrics. However, this projection is typically approximate, as observed distances derived from data rarely lie exactly in tree space. 

While these classical methods often produce high-quality trees, they incur substantial computational cost for each inference due to the combinatorial nature of tree search. More recently, deep learning approaches have been proposed to infer phylogenetic structure directly from MSAs \citep{blassel2025likelihood}. These methods shift the computational burden to a training phase, after which inference can be performed rapidly by a forward pass through a trained model. This approach offers a scalable alternative to traditional per-instance optimization. 

One representative example is Phyloformer which adopts a transformer-based architecture operating on multiple sequence alignments (MSAs) \citep{nesterenko2025phyloformer}. The model encodes sequences into pairwise representations and predicts evolutionary distances, which are then used to reconstruct trees. By replacing explicit likelihood computation with learned representations, Phyloformer achieves competitive or improved accuracy relative to classical distance-based and likelihood-based methods while significantly reducing inference time per instance \citep{nesterenko2025phyloformer}. However, its performance depends on the distribution of training data, and like many supervised models, it can degrade under shifts in sequence length, evolutionary rate, or substitution model.

A more recent line of work extends this paradigm by directly modeling distributions over tree topologies. For example, the Phyloformer 2 uses likelihood-free framework to construct a neural parameterization of the posterior over phylogenies \citep{blassel2025likelihood}. The architecture uses a simulation-based approach called Neural Posterior Estimation (NPE) to estimate evolutionary tree distributions without computing complex likelihood functions. The model uses normalizing flows to map summary statistics from observed and simulated sequence data to infer posterior distributions, from which they directly infer trees. The model demonstrates improved robustness to model misspecification, suggesting stronger generalization compared to classical methods that rely on restrictive assumptions. Phyloformer 2 typically outperforms Phyloformer topologically but underperforms on branch-length aware metrics \citep{blassel2025likelihood}.

Despite these advances, both architectures implicitly learn representations in an ambient Euclidean space. As a result, these models produce outputs that violate tree metric constraints or require post hoc correction. Further, although likelihood-free and neural approaches improve computational efficiency and can outperform classical methods they struggle to generalize to MSAs out-of-distribution (OOD) from the training data which is not an issue for classical methods \citep{blassel2025likelihood, nesterenko2025phyloformer}.

\section{Tropical Deep Learning}

Our method is motivated by the mismatch between the geometry of phylogenetic tree space and the Euclidean geometry implicitly assumed by standard neural architectures. Although pairwise distances can be embedded in $\mathbb{R}^{\binom{n}{2}}$, valid tree metrics occupy a structured, non-Euclidean subset shaped by the four point condition. To address this mismatch, we adopt \emph{tropical attention} introduced by \citep{hashemi2025tropical}, where comparisons are governed by max-plus structure and the resulting maps are piecewise-linear and scale-invariant in the projective sense. This aligns the model's inductive bias with the geometry of phylogenetic inference since the target objects are themselves governed by combinatorial and hierarchical structure, not arbitrary linear variation. This geometric reality also motivate a linear combination of loss functions that reflect aspects of the tree inference problem: $\ell_1$ for coordinate accuracy, tropical symmetric loss to penalize distortions from target tropical geometry, and ultrametric loss for violations of the rooted tree constraints (and implicitly unrooted tree constraints). Thus, both the network mechanism and the training objective are chosen to reflect the underlying geometry of tree inference.
\subsection{Loss Functions}

Our objective combines ambient accuracy with geometry-aware regularization. Let $\hat d \in \mathbb{R}^{\binom{n}{2}}$
denote the predicted vector of pairwise leaf distances and let $d \in \mathbb{R}^{\binom{n}{2}}$ denote the ground-truth distance vector. Although both objects are represented in the Euclidean ambient space \(\mathbb{R}^{\binom{n}{2}}\), the subset corresponding to valid tree metrics is highly structured and is better understood through tropical and tree constraints than through ordinary Euclidean geometry alone. For this reason, we do not train solely with a regression loss as in the Phyloformer model \citep{nesterenko2025phyloformer}. Instead, we combine an \(\ell_1\) term, which measures coordinatewise accuracy, with a tropical symmetric term, $d_{tr}$, which measures distortion in tropical geometry, and an ultrametric term, which biases predictions toward the rooted tree regime of interest.

The first component is the \(\ell_1\) loss,
\[
L_{1}(\hat d,d)=\|\hat d-d\|_{1},
\]
which directly penalizes absolute prediction error in the ambient vector space. This term stabilizes optimization and ensures that the model learns outputs close to the observed targets. However, by itself, \(\ell_1\) treats \(\hat d\) and \(d\) as arbitrary Euclidean vectors and does not account for the projective, polyhedral structure that distinguishes tree-valued outputs from generic points in \(\mathbb{R}^{\binom{n}{2}}\). 
In our setting, this yields the tropical symmetric loss
\[
d_{tr}(\hat d,d)
=
\max_{i}(\hat d_{i}-d_{i})
-
\min_{i}(\hat d_{i}-d_{i}).
\]
The tropical metric has been widely used as a natural metric on tree spaces \citep{monod2022tropical}. Because of its projective nature, this metric depends on relative discrepancies rather than absolute differences, and it encourages the network to preserve the combinatorial shape of the target distance pattern rather than merely minimizing coordinatewise error. 

To further bias the output toward phylogenetic tree structure, we add an ultrametric penalty. For each triple of leaves $(i,j,k)$, we extract the three predicted distances
\[
\hat d_{ij},\quad \hat d_{ik},\quad \hat d_{jk},
\]
identify the two largest values, and penalize any gap between them. Writing the ordered top two values as \(a \geq b\), the triplewise violation is
\[
\max(a-b,\,0).
\]
The ultrametric loss \(L_{\mathrm{ultra}}(\hat d)\) is then defined as the average of this quantity over all triples. Thus, the penalty is zero exactly when, for every triple, the two largest distances are equal, and increases as triples deviate from ultrametric structure. In this way, the loss provides a differentiable measure of how far the prediction lies from $\mathcal{U}_n$, the space of ultrametrics.

Equation \ref{losses} reflects a practical training paradigm to leverage the benefits discussed previously. During the first 20 epochs, optimization focuses on recovering the coarse distance geometry of the target through Euclidean accuracy and tropical consistency. Once the model has learned a reasonable representation of pairwise relationships, the ultrametric penalty is activated to encourage convergence toward the rooted-tree subspace without constraining early learning.
\begin{equation}\label{losses}
    \mathcal{L}_e=
\begin{cases}
1.0\,L_1(\hat d,d)+1.4\,d_{tr}(\hat d,d), & e \leq 20,\\[4pt]
1.0\,L_1(\hat d,d)+1.4\,d_{tr}(\hat d,d)+0.6\,L_{\mathrm{ultra}}(\hat d), & e>20.
\end{cases}
\end{equation}

\subsection{Tropical Attention}

The central contribution of this model is to replace vanilla softmax attention with the algorithmically aligned tropical attention mechanism. Vanilla softmax dot-product attention operates in smooth Euclidean space and converts logits into dense probability distributions which tend to spread strictly positive mass across all tokens. In problems whose correct solution depends on decisive selection or comparison, this smoothing can blur the sharp boundaries that define the underlying computation. \cite{hashemi2025tropical} argue that this is a fundamental source of failure for softmax-based reasoning on combinatorial tasks, especially under out-of-distribution changes in sequence length or value scale, because the top logit gaps can shrink and the resulting attention becomes increasingly diffuse. In contrast, many combinatorial and dynamic-programming computations are naturally governed by max, min, and additive operations, which are piecewise-linear and polyhedral rather than smooth \citep{jukna2014lower}.

Tropical attention is designed to build this inductive bias directly into the reasoning mechanism. Following \citep{hashemi2025tropical}, the input representation is first mapped into tropical projective space through a learnable valuation map, producing a tropicalized representation. Queries, keys, and values are then formed by tropical linear projections in the max-plus semiring. 
{Specifically, let $R\in \mathbb{R}^{M\times P \times d_{model}}$ be the token embedding of the input. Then here we define the \emph{tropicalization map} by going to an amoeba representation of the input with a learnable valuation map, $\Phi: \mathbb{R}^{M\times P \times d_{model}}\rightarrow (\mathbb{TP}^{d_{model}-1})^{M\times P}$ such that 
\begin{equation}
\Phi_\lambda(R)_{ij}\;=\;
\mathbf{X}_{ij}- \max_{1\leq r\leq d_{model}}\mathbf{X}_{ijr}.\mathbf{1}_{d_{model}},~~\text{where}~~\mathbf{X} = \log\bigl(\max({\bf{0}}, R)\bigr) \in \mathbb{R}^{M\times P \times d_{model}}
\label{eq:phi}
\end{equation}
for each row $i\in \{1,\dots, M\}$ and for each column $j 
{\in} \{1,\dots, P\}$. The constant shift enforces $\max_{i j}\phi_\lambda(\mathbf x)_{ij} = \epsilon$, so the output of $\phi_\lambda$ always lies in the tropical simplex, $\Delta^{\,d_{model}-1}:= \bigl\{\,z\in\mathbb R^{d_{model}}\big|\max_{ij}z_{ij} = \epsilon \bigr\}$, where every vector is projectively equivalent to exactly one point in the tropical simplex.}

For the tropicalized input $X$, the projections 
{
\begin{equation}
{Q = \Phi_\lambda(R)\odot W^Q, \qquad K = \Phi_\lambda(R)\odot W^K, \qquad V = \Phi_\lambda(R)\odot W^V}
\end{equation}
}
are computed by replacing the linear map with tropical matrix multiplication. Instead of computing similarity by Euclidean dot product and normalizing with softmax, tropical attention scores query-key agreement using the tropical metric,
{
\begin{equation}\label{eq:d_tr_qk}
    d_{tr}({Q, K})
= S_{i,j} = - ( 
\max({q_i-k_j})-\min({q_i-k_j})).
\end{equation}
}
The attention scores, $S$, are then aggregated tropically with $V$ step via tropical matrix multiplication
{
\begin{equation}\label{eq:attn_scores}
    \textit{Attn}(Q,K,V)_i = \max_j(S_{i,j}+v_j).
\end{equation}
}
Conceptually, this replaces the smooth weighting rule “average all values with exponentially reweighted coefficients” by the sharp rule “select the value that aligns best with the query in the tropical projective torus.” The output is then mapped back into ordinary Euclidean coordinates so that the remainder of the network can can be trained via traditional deep learning gradient descent methods and existing packages \citep{hashemi2025tropical}.

This construction changes the geometry of the computation. In the tropical-attention framework, the {computation is} piecewise-linear {and uses} idempotent {max-plus} aggregation, {while its constituent tropical linear maps are} non-expansive in the tropical metric \citep[Proposition 9]{talbut2024probabilitymetricstropicalspaces}. This preserves the polyhedral decision structure characteristic of combinatorial value functions, rather than smoothing those boundaries into quadratic or spherical regions as softmax attention does. 

This inductive bias is attractive for phylogenetic inference even though the task is not identical to the combinatorial benchmarks studied in the paper. Tree metrics, ultrametrics, and tropical tree spaces are governed by polyhedral structure and constraints \citep{monod2022tropical}. Smooth softmax attention is therefore not the most natural mechanism for propagating information relevant to hierarchical tree structure. Tropical attention encourages sharp relational comparisons and projectively meaningful aggregation, which is better aligned with distance-based tree prediction. In this sense, our use of tropical attention is not merely a novel attention trick, it is a geometric choice intended to match the architecture’s inductive bias to the geometry of the target objects. 

\section{Model}
Our model is an extension of both the tropical attention model proposed by \cite{hashemi2025tropical} and the Phyloformer model introduced by \cite{nesterenko2025phyloformer}, using \emph{tropical axial attention} on MSAs to predict distance matrices. The overall design closely matches that of Phyloformer, and can be broken into three stages: data embedding, tropical axial attention, and prediction.

MSA input data is a set of $n$ taxa, which will be called leaves for the remainder of the section. Each leaf is a sequence of observed residues of length $M$. Each residue is from an alphabet of size $A$. The sequence is then one-hot encoded and embedded in dimension $d_{model}$. Finally, each unordered pair of leaves is concatenated and re-projected to dimension $d_{model}$. The full embedding is as follows,
\begin{equation}
    \bigg{[} M, n \bigg] \rightarrow \bigg[M, n, A\bigg] \rightarrow \bigg[M,n,d_{model}\bigg] \rightarrow \bigg[M, \binom{n}{2}, 2d_{model}\bigg] \rightarrow \bigg[M, \binom{n}{2}, d_{model}\bigg].
\end{equation}
After this step, each token is then a representation of an unordered leaf pair at a particular alignment position and $T = [M] \times \mathcal{P}$ where $\mathcal P = \{(i,j): 1 \leq i {<} j \leq n{\}}$. For the remainder of the section, we will denote the unordered pairs $P=\binom{n}{2}$ and the embedded tensor $R \in \mathbb{R}^{M\times P \times d_{model}}$. Each token $r_{m,(i,j)} \in \mathbb{R}^{d_{model}}$ contains site specific information about the evolutionary relationship between leaves $i$ and $j$. Evaluating pairwise vice simply leaf-wise information is motivated by the structure of the model objective. As explained previously, the  tree pairwise distance matrix is distinguished from that of a general connected graph by its hierarchical structure and compliance with four point and ultrametric conditions, which are constraints on sets of pairwise distances. This property means the target is constrained by relationships \emph{between} leaves rather than the properties of individual leaves.

\begin{figure}
    \centering
    \includegraphics[width=0.75\linewidth]{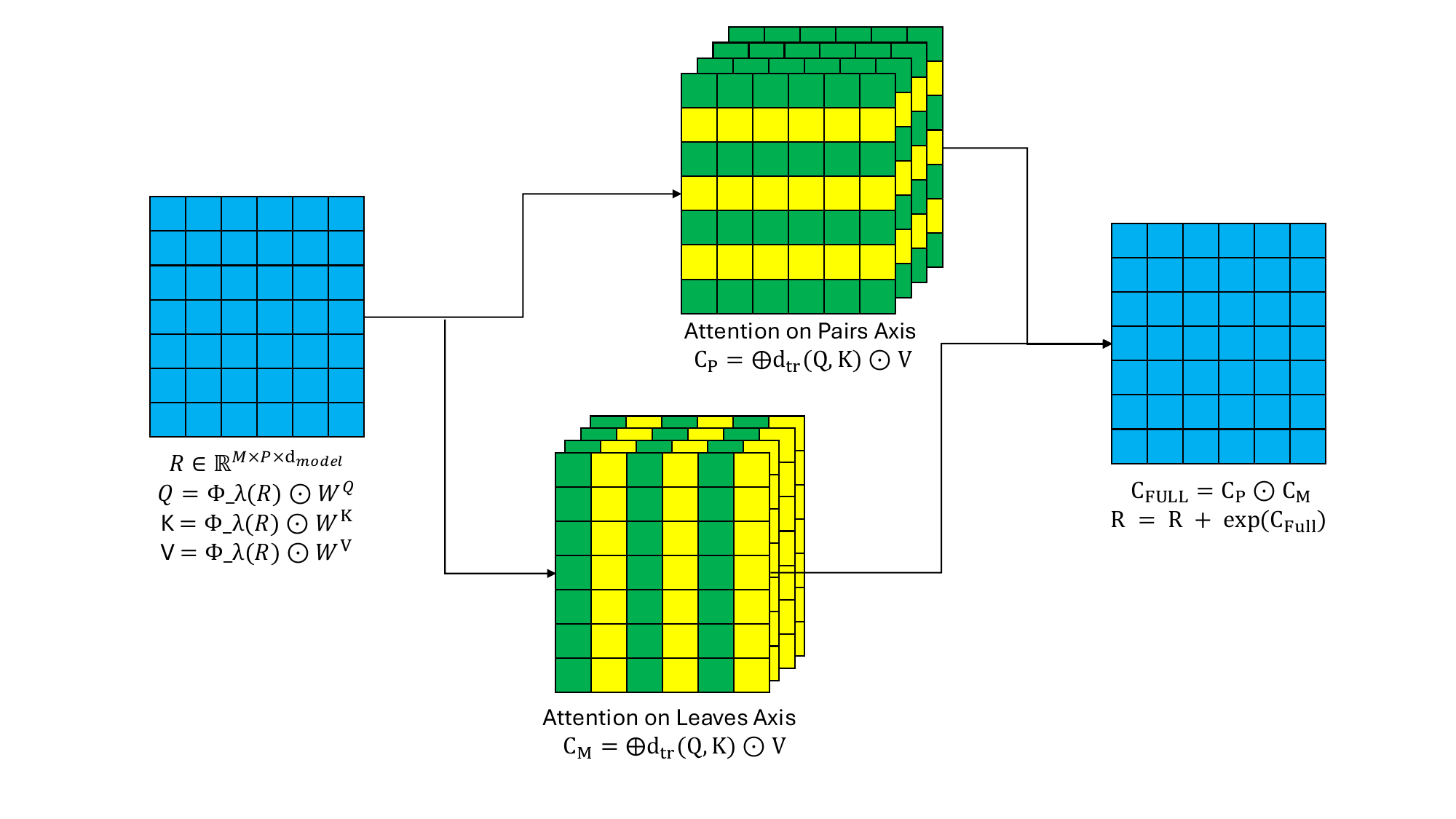}
    \caption[Tropical Axial Attention]{Tropical axial attention takes $R\in \mathbb{R}^{M\times P \times d_{model}}$ and tropically projects to $Q, K, V$ tensors. Attention is performed in parallel along both the $M$ and $P$ axis, preserving the geometric relevance of both axes for phylogenetic inference.}
    \label{fig:axialattention}
\end{figure}

The embedded tensor $R \in \mathbb{R}^{M \times P \times d_{model}}$ is then input into the main portion of the model which consists of six tropical axial attention blocks of four heads each. Due to the multi-axial nature of the data - sequence positions $M$ and leaf pairs $P$ - we implement axial attention from \cite{lucidrains2021axial} to attend to both axes \citep{ho2019axial, wang2020axialdeeplab}, see Figure \ref{fig:axialattention}. Each block alternates attention over these axes so that information propagates across both sequence positions and pairwise relationships, preserving the biologically and geometrically relevant information \cite{nesterenko2025phyloformer}. 
\begin{itemize}
    \item \emph{Sequence axis attention}: For a fixed leaf pair $(i,j) \in P$, attention across $M$ sites lets the model learn mutation evidence across the alignment.
    \item \emph{Pair axis attention}: For a fixed site $m \in [M]$, attention across pairs $P$ computes relationships between pairs which are the elements of geometrically relevant tree-constraints which distinguish tree metrics from ordinary distance matrices, see Theorems \ref{thm:fourpoint} , \ref{thm:ultrametric}.
\end{itemize}
In our model, the vanilla softmax dot-product attention is replaced with tropical attention introduced in the previous section. Thus, queries, keys, and values are computed through tropical linear projections, compared through the projectively invariant tropical metric, and aggregated by tropical matrix multiplication rather than convex Euclidean operations. This gives each block the ability to refine pairwise phylogenetic representations while preserving the polyhedral, projectively meaningful comparisons that define the tree inference problem. 

After the attention stack, the resulting $\big[M, P, d_{model}\big]$ tensor is mapped to a scalar prediction for each pair of leaves with a pointwise feed-forward projection via $(1\times 1)$ convolution. This convolution is a channel-wise projection of each sequence position/pair location, $(m,(i,j)) \in T$ from $\mathbb{R}^{d_{model}} \to \mathbb{R}$. Because the projection is applied independently at each $(m, (i,j))$ location, it collapses only the channel dimension while preserving the learned sequence- and pair-dimension information. Softplus then ensures non-negative outputs appropriate for distance prediction, and the mean is taken across the sequence dimension to yield the pairwise distance prediction $\hat y \in \mathbb{R}^P_{\geq 0}$.

 With a pairwise axial structure and the projectively invariant similarity measure of tropical attention, our model closely aligns with the basic relational nature of the phylogenetic tree inference problem. The full architecture combines the biologically sound embedding pipeline of Phyloformer \citep{nesterenko2025phyloformer} with the more geometrically aligned inductive bias of tropical attention, designed to improve neural phylogenetic inference. 

\section{Experiments}
In this section, we evaluate our proposed tropical axial attention model against both the Phyloformer of \cite{nesterenko2025phyloformer} and Phyloformer 2 of \cite{blassel2025likelihood}. The experimental objective is to determine if replacing Euclidean softmax dot-product attention with tropical axial attention improves phylogenetic distance estimation and aides tree reconstruction, particularly when evaluating data from outside the training data distribution.
{In this experiments, we use the FastME implementation \citep{lefort2015fastme}, a heuristic for the BME, to {infer a phylogenetic tree from } a distance matrix.}

\subsection{Experimental Design}
To isolate architectural differences, we train all models on the same set of $10,000$ simulated MSAs with $n=10$ leaves and sequence length $M=250$. The training data are simulated utilizing the same birth-death process under the Alisim LG+GC model \citep{LyTrong2022AliSim} used in both Phyloformer and Phyloformer 2 \citep{nesterenko2025phyloformer, blassel2025likelihood}. The comparison models are trained using the hyperparameter settings described in their respective papers, except we use a standard embedding dimension of $d_{model}=80$ for all models to control their representational capacity. We train the tropical model with the same $L = 6$ axial attention blocks of $H=4$ heads each as the Phyloformer model, and we use a learning rate of $\lambda=0.001$. {These hyperparameters product a tropical model with 636,001 trainable parameters and Phyloformer model with 477,697 trainable parameters.} We train each model on the same set of $5$ random seeds and report performance as an average across seeds.

All models follow the same input-output structure, receiving an input MSA and outputting a vector of predicted pairwise leaf distances. These predicted distances are converted into a distance matrix which is then used to infer a tree for topological comparison using FastME \citep{lefort2015fastme}. This design allows comparison of models both of distance estimators and as components of a full phylogenetic inference pipeline.

\subsection{Evaluation Metrics}
We evaluate the tree inference performance of all models based on topological accuracy by comparing the predicted tree to the true reference tree. The primary topological metric is the Robinson-Foulds (RF) distance \citep{Robinson1981ComparisonOP}. Let $\hat G$ and $G$ be the predicted and true trees, respectively, and $\mathcal{S}(G)$ be the set of bipartitions of the internal edges. The RF distance \[d_{RF}\big(\hat G, G\big) = \big|\mathcal{S}(G)\setminus \mathcal{S}(\hat G)\big| + \big|\mathcal{S}(\hat G)\setminus \mathcal{S}(G)\big|,\] where $|\cdot |$ denotes the cardinality of the set. The RF distance exclusively compares the topological difference between $G$ and $\hat G$.

We also consider extensions of the RF distance. The first normalizes by the total number of possible splits on fully resolved unrooted binary trees with $n$ leaves, yielding the normalized RF distance \[d_{nRF} = \frac{d_{RF}}{2n -6},\] which controls for the size of the tree \citep{Robinson1981ComparisonOP}. 

The second RF extension, weighted RF distance \[d_{wRF}(\hat G, G) = \sum_{\sigma \in \mathcal{S}(\hat G) \cup \mathcal{S}(G)} \big|w_1(\sigma)-w_2(\sigma)\big|,\] where $\sigma$ is a split in either tree, $w_{\hat G}(\sigma)$ is the branch length of that split in $\hat G$,  $w_G(\sigma)$ is the branch length of the split in $G$, and $w_i(\sigma) = 0$ if $\sigma$ is not in tree $i$. Thus, the weighted RF distance incorporates both the topology and branch length deviation between $G$ and $\hat G$ \citep{Robinson1981ComparisonOP}.

The final tree error distance is the Kuhner-Felsenstein (KF) distance \[d_{KF}(\hat G, G) = \sqrt{\sum_{\sigma \in \mathcal{S}(\hat G) \cup \mathcal{S}(G)} ( w_1(\sigma)-w_2(\sigma))^2}.\] Like weighted RF, KF penalizes both topological and branch-length differences, but uses an $\ell_2$ rather than $\ell_1$ penalty \citep{kuhner1994simulation}.

The distances above each evaluate how close an inferred tree is from the simulated trees associated with the input MSA, but we are also interested in geometric performance and how well each model learns outputs in tree space. To do so, we first infer a valid tree $\hat G \in \mathcal{T}_n$ from $\hat d$ with the FastME implementation \citep{lefort2015fastme}. We then calculate the pairwise path lengths for $\hat G$ which provides a valid pairwise distance vector $\pi_{{FastME}}(\hat d) \in \mathcal{T}_n${, which is obtained via FastME from $\hat d$ (in order to simplify, we denote as $\pi_{FastME}$ thereafter)}. Finally, we evaluate $d_{tr}(\hat d, \pi_{{FastME}})$ and $\text{MAE}(\hat d, \pi_{{FastME}}) = \frac{1}{P}\sum_{1\leq i < j \leq n} \Vert \hat d_{ij} - \text{proj}(\hat d_{ij}) \Vert_1$ to measure how far $\hat d$ is from $\mathcal{T}_n$ both projectively and absolutely. This assesses the consistency of each model as an estimator for valid tree metrics.

\subsection{Results}\label{sec:phy_results}
\begin{figure}[ht]
    \centering
    \includegraphics[width=\textwidth]{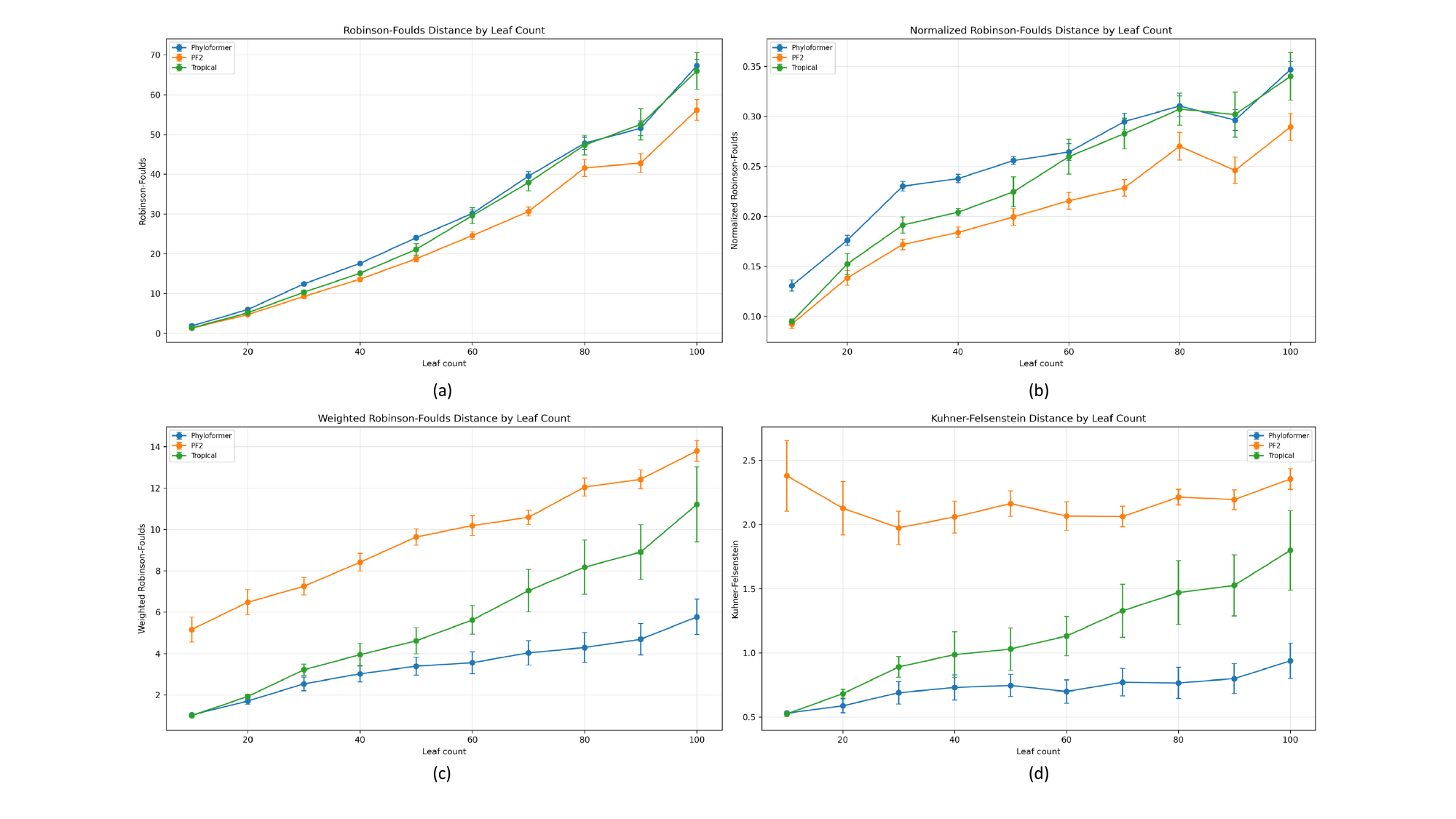}
    \caption[Tree Reconstruction Performance]{Tree reconstruction error by leaf count on simulated test data generated under the training regime. Panels show (a) RF distance, (b) normalized RF distance, (c) weighted RF distance, and (d) KF distance. Lower values indicate better agreement with the reference tree associated with the input MSA. Phyloformer 2 achieves the strongest topological performance under RF-based metrics, while Phyloformer generally has the lowest branch-length-sensitive error under weighted RF and KF distances. The Tropical model performs generally between the other two models on both topological and branch-length sensitive metrics.}
    \label{fig:test_tree_metrics_by_leaf_count}
\end{figure}

We evaluate model performance under a variety of input distributions. The first experiment evaluates performance on simulated data generated in the same manner as the training set with $250$ MSAs for each of $n \in \{10,20,30,40,50\}$ and $50$ MSAs for each of $n \in \{50,60,70,80,90\}$. All MSAs are generated with sequence length $M=200$. Results of this experiment are shown in Figure \ref{fig:test_tree_metrics_by_leaf_count} and Table \ref{tab:best_checkpoint_across_tips}. {We used the Wilcoxon signed-rank test to perform 
pairwise comparisons between models for each metric
\citep{wilcoxon1945individual,hollander2013nonparametric},  
where each test instance constituted an 
observation. Because multiple pairwise comparisons
were conducted within each metric and tree size,
adjusted p-values were computed using Holm's
sequential multiple-testing procedure
\citep{holm1979simple}. 
}
\begin{table}[ht]
  \centering
  \caption{Pairwise comparison of Tropical with baseline models, averaged across tree sizes. Each tree constitutes one paired observation in the Wilcoxon signed-rank test. The reported model values are arithmetic means across tree sizes. The difference $\Delta$ is Tropical minus the baseline, so negative values show Tropical outperformance.}
  \label{tab:best_checkpoint_across_tips}
  \begin{tabular}{llrrrrl}
    \toprule
    Metric & Comparison & Tropical & Baseline & $\Delta$ & $p$-value \\
    \midrule
    RF & Tropical vs. PF & 28.635 & 29.813 & -1.178 & 0.020 \\
     & Tropical vs. PF2 & 28.635 & 24.338 & +4.297 & 0.004 \\
    Normalized RF & Tropical vs. PF & 0.2359 & 0.2543 & -0.0184 & 0.010 \\
     & Tropical vs. PF2 & 0.2359 & 0.2036 & +0.0323 & 0.004 \\
    Weighted RF & Tropical vs. PF & 5.565 & 3.402 & +2.164 & 0.004 \\
     & Tropical vs. PF2 & 5.565 & 9.599 & -4.033 & 0.004 \\
    KF & Tropical vs. PF & 1.135 & 0.724 & +0.411 & 0.004 \\
     & Tropical vs. PF2 & 1.135 & 2.158 & -1.023 & 0.004 \\
    \bottomrule
  \end{tabular}
\end{table}

These experiments are designed to evaluate the input size generalization capability of each model. The $n=10$ test samples represent in-distribution performance while the $n=20$ through $n=100$ samples represent increasing length OOD performance. Since the number of pairwise distances grows with $\binom{n}{2}$, the experiment tests how well each model can represent coherent structure as the output dimension grows beyond training inputs. Consistent with the results in \citep{blassel2025likelihood}, {we see from Table \ref{tab:best_checkpoint_across_tips} } that the Phyloformer 2 model shows consistently superior topological performance while Phyloformer outperforms on metrics that consider branch lengths. The tropical model consistently achieves results between those of Phyloformer and Phyloformer 2 across all metrics. Additionally, the tropical model appears to match the performance degradation trend of both baseline models on the topologically focused RF-based metrics as number of leaves increases, but performance degrades on the branch-length aware distances.

The second experiment evaluates model performance on simulated data with a different generation process. This data set is $30$ $40$-taxon MSAs and $10$ $100$-taxon MSAs both with $M=500$ length sequences and corresponding true trees from the PhyML benchmark \citep{guindon2010new}. In contrast to the birth-death process used during training, this benchmark was generated using the simulation design of \cite{guindon2003simple}, which generated $40$-taxon trees using the process of \cite{kuhner1994simulation}. Alignments were simulated under a General Time Reversible (GTR) model with $4$ gamma-distributed rates \citep{lanave1984new, guindon2010new}. This benchmark provides a useful test of the ability of each model to generalize to data from unseen distributions.

\begin{table}[ht]
\centering
\caption[PhyML Tree Statistics]{PhyML \citep{guindon2010new} tree statistics for Phyloformer (PF), Phyloformer 2 (PF2), and Tropical (Trop) models on $n=40, 100$ leaves. Values are mean $\pm$ standard deviation.}
\label{tab:phyml_treestats_best}
\begin{tabular}{llcccc}
\toprule
\textbf{Model Type} & \textbf{Leaf Count} & \textbf{RF} & \textbf{Norm. RF} & \textbf{Weighted RF} & \textbf{KF Score} \\
\midrule
PF        & 40  & $22.96 \pm 8.37$  & $0.310 \pm 0.113$ & $1.31 \pm 0.74$  & $0.287 \pm 0.162$ \\
PF2       & 40  & $7.95 \pm 0.20$   & $0.107 \pm 0.003$ & $13.00 \pm 1.00$ & $2.939 \pm 0.267$ \\
Trop & 40  & $9.65 \pm 1.73$   & $0.130 \pm 0.023$ & $0.71 \pm 0.38$  & $0.165 \pm 0.083$ \\
\hline
PF        & 100 & $57.44 \pm 26.73$ & $0.296 \pm 0.138$ & $3.52 \pm 1.73$  & $0.481 \pm 0.256$ \\
PF2       & 100 & $25.24 \pm 5.23$  & $0.130 \pm 0.027$ & $20.33 \pm 2.29$ & $3.108 \pm 0.373$ \\
Trop & 100 & $31.00 \pm 4.66$  & $0.160 \pm 0.024$ & $2.25 \pm 1.06$  & $0.320 \pm 0.146$ \\
\bottomrule
\end{tabular}
\end{table}
Tables \ref{tab:phyml_treestats_best} and \ref{tab:phyml_distancestats_best} show results on the PhyML dataset. In contrast to the birth-death generated data, we see that the tropical model performs comparably with Phyloformer 2 topologically while outperforming both other models when considering branch lengths. Table \ref{tab:phyml_distancestats_best} shows that the tropical model is learning distance matrices closer to tree-metric distances than the other methods. Taken together, we see the tropical model achieves better relative performance on the OOD PhyML dataset than in the test data generated with the same birth-death process as the training data. This indicates distributional robustness and supports the hypothesis that the tropical model learns structure which generalizes beyond the training simulation regime.
\begin{table}[ht]
\centering
\caption[PhyML Distance Statistics]{PhyML \citep{guindon2010new} distance statistics for Phyloformer (PF), Phyloformer 2 (PF2), and Tropical (Trop) models on $n=40, 100$ leaves. Values are mean $\pm$ standard deviation. Distances were calculated between $\hat d \in \mathbb{R}^{\binom{n}{2}}$ and $\pi_{{FastME}} \in \mathcal{T}_n$.}
\label{tab:phyml_distancestats_best}
\begin{tabular}{llcc}
\toprule
\textbf{Model Type} & \textbf{Leaf Count} & \textbf{MAE} & \textbf{Tropical Distance} \\
\midrule
PF & 40  & $0.0265 \pm 0.0053$ & $0.2843 \pm 0.0784$ \\
PF2         & 40  & $3.9345 \pm 0.2653$ & $5.9005 \pm 0.3548$ \\
Trop   & 40  & $0.0047 \pm 0.0030$ & $0.0416 \pm 0.0260$ \\
\hline
PF & 100 & $0.0496 \pm 0.0044$ & $0.5909 \pm 0.1243$ \\
PF2         & 100 & $3.9226 \pm 0.3827$ & $6.0401 \pm 0.4555$ \\
Trop    & 100 & $0.0070 \pm 0.0051$ & $0.0709 \pm 0.0514$ \\
\bottomrule
\end{tabular}
\end{table}

The final experiment consists of empirical datasets. We use $11$ MSAs, known as DS1-DS11, which serve as a benchmark for empirical evaluation~\citep{whidden2015quantifying}. Summary details for these data sets are outlined in Table \ref{tab:empirical_datasets}, and full details can be found in \citep{lakner2008efficiency}.

\begin{table}[H]
\centering
\caption[Empirical Phylogenetic Data]{Empirical DS1-DS11 datasets used for phylogenetic evaluation \citep{lakner2008efficiency, whidden2015quantifying}. Each data set represents a different type of data of interest to the phylogenetic community following different evolutionary patterns.}
\label{tab:empirical_datasets}
\begin{tabular}{lrrlp{2.5cm}r}
\toprule
Data & $N$ & Cols & Type of Data  & Est. Error \\
\midrule
DS1  & 27 & 1949 & rRNA; 18S  & 0.0048 \\
DS2  & 29 & 2520 & rDNA; 18S  & 0.0002 \\
DS3  & 36 & 1812 & mtDNA; COII (1678); cytb (679--1812) & 0.0002 \\
DS4  & 41 & 1137 & rDNA; 18S & 0.0006 \\
DS5  & 50 & 378  & Nuclear protein coding; wingless & 0.0005 \\
DS6  & 50 & 1133 & rDNA; 18S  & 0.0023 \\
DS7  & 59 & 1824 & mtDNA; COII; and cytb & 0.0011 \\
DS8  & 64 & 1008 & rDNA; 28S &  0.0009 \\
DS9  & 67 & 955  & Plastid ribosomal protein; s16 (rps16) & 0.0164 \\
DS10 & 67 & 1098 & rDNA; 18S & 0.0164 \\
DS11 & 71 & 1082 & rDNA; internal transcribed spacer & 0.0008 \\
\bottomrule
\end{tabular}
\end{table}

Empirical inference is particularly important in the phylogenetic setting because real-world data includes biological structure not captured under any simulation regime, and performance on this data indicates whether the model is simply fitting the training simulation prior or learning the structure of tree metrics. Unlike the simulated experiments, the true trees of empirical MSAs are unknown. Therefore, this experiment tests model output behavior on real sequence data rather than on an ability to reconstruct a synthetic tree. 

\begin{table}[ht]
\centering
\caption[Empirical DS1-DS11 Distance Statistics]{Empirical matrix-to-tree consistency distance metrics for Phyloformer (PF), Phyloformer 2 (PF2), and Tropical (Trop) models. Values are mean $\pm$ standard deviation across seeds. Distances were calculated between $\hat d \in \mathbb{R}^{\binom{n}{2}}$ and $\pi_{{FastME}} \in \mathcal{T}_n$. Lower values indicate greater consistency between the predicted matrix and its induced tree representation.}
\vskip 0.2in

\label{tab:empirical_matrix_tree_consistency_by_sample}
\begin{tabular}{llcc}
\hline
\textbf{Sample} & \textbf{Model} & \textbf{MAE} & \textbf{Tropical Distance} \\
\hline
DS1 & PF & $0.0356 \pm 0.0170$ & $0.3177 \pm 0.1203$ \\
DS1 & PF2 & $0.7434 \pm 0.3203$ & $5.4046 \pm 1.8920$ \\
DS1 & Trop & $0.0066 \pm 0.0035$ & $0.0555 \pm 0.0316$ \\
\hline
DS2 & PF & $0.0280 \pm 0.0147$ & $0.2753 \pm 0.1180$ \\
DS2 & PF2 & $0.2606 \pm 0.0246$ & $2.4208 \pm 0.2262$ \\
DS2 & Trop & $0.0063 \pm 0.0033$ & $0.0528 \pm 0.0273$ \\
\hline
DS3 & PF & $0.0319 \pm 0.0196$ & $0.3147 \pm 0.1972$ \\
DS3 & PF2 & $0.2204 \pm 0.0184$ & $2.1074 \pm 0.1991$ \\
DS3 & Trop & $0.0036 \pm 0.0020$ & $0.0332 \pm 0.0179$ \\
\hline
DS4 & PF & $0.0663 \pm 0.0238$ & $0.8848 \pm 0.5287$ \\
DS4 & PF2 & $0.8826 \pm 0.2372$ & $7.1742 \pm 1.5143$ \\
DS4 & Trop & $0.0104 \pm 0.0047$ & $0.0965 \pm 0.0414$ \\
\hline
DS5 & PF & $0.0522 \pm 0.0333$ & $0.4895 \pm 0.3107$ \\
DS5 & PF2 & $0.5476 \pm 0.0539$ & $5.0285 \pm 0.4198$ \\
DS5 & Trop & $0.0106 \pm 0.0065$ & $0.1029 \pm 0.0597$ \\
\hline
DS6 & PF & $0.0426 \pm 0.0128$ & $0.4344 \pm 0.1397$ \\
DS6 & PF2 & $0.8515 \pm 0.1983$ & $6.5934 \pm 1.6151$ \\
DS6 & Trop & $0.0072 \pm 0.0027$ & $0.0689 \pm 0.0304$ \\
\hline
DS7 & PF & $0.0271 \pm 0.0123$ & $0.2758 \pm 0.1349$ \\
DS7 & PF2 & $0.2198 \pm 0.0173$ & $2.2003 \pm 0.1822$ \\
DS7 & Trop & $0.0041 \pm 0.0024$ & $0.0387 \pm 0.0217$ \\
\hline
DS8 & PF & $0.1602 \pm 0.0750$ & $1.4613 \pm 0.9390$ \\
DS8 & PF2 & $0.9784 \pm 0.6253$ & $5.6614 \pm 1.8840$ \\
DS8 & Trop & $0.0146 \pm 0.0087$ & $0.1217 \pm 0.0516$ \\
\hline
DS9 & PF & $0.0166 \pm 0.0189$ & $0.4166 \pm 0.3514$ \\
DS9 & PF2 & $0.2760 \pm 0.0461$ & $4.9927 \pm 0.6803$ \\
DS9 & Trop & $0.0022 \pm 0.0012$ & $0.0567 \pm 0.0292$ \\
\hline
DS10 & PF & $0.0262 \pm 0.0187$ & $0.3007 \pm 0.2047$ \\
DS10 & PF2 & $1.1232 \pm 0.1705$ & $7.8954 \pm 1.2276$ \\
DS10 & Trop & $0.0068 \pm 0.0058$ & $0.0651 \pm 0.0385$ \\
\hline
DS11 & PF & $0.0218 \pm 0.0132$ & $0.5941 \pm 0.3022$ \\
DS11 & PF2 & $0.1808 \pm 0.0286$ & $2.4901 \pm 0.4258$ \\
DS11 & Trop & $0.0075 \pm 0.0053$ & $0.0740 \pm 0.0427$ \\
\hline
\end{tabular}
\end{table}

Table \ref{tab:empirical_matrix_tree_consistency_by_sample} shows matrix-tree space results for the DS1-DS11 datasets. We see that the tropical model is much closer both projectively and absolutely to its {FastME} reconstruction than either of the other models with lower variance. This suggests, even when presented with empirical data outside of the training distribution, the tropical model provides a more consistent estimator for tree metrics than Phyloformer and Phyloformer 2. 

\subsection{Ablation Study}

{To isolate the contributions of the attention mechanism and individual loss terms, we conducted an ablation study using the final six-block, four-head architecture. Each configuration was trained on the same 1,000 simulated $10$-taxon alignments using five random seeds and evaluated on the same 200 held-out $10$-taxon alignments. We compared tropical attention trained with $\ell_1$ loss only, tropical loss only, $\ell_1$ plus tropical loss, $\ell_1$ plus tropical loss plus the ultrametric penalty, and a softmax attention model of the same construction with just the tropical loss. Table \ref{tab:ablation} reports mean $\pm$ standard deviation across seeds.}

\begin{table}[t]
\centering
\small
\setlength{\tabcolsep}{4pt}

\begin{tabular}{@{}lcccc@{}}
\toprule
Configuration
& RF
& Norm. RF
& Weighted RF
& KF \\
\midrule
Trop. + $\ell_1$
& $13.3470 \pm 0.0268$
& $0.9516 \pm 0.0019$
& $1.8428 \pm 0.0022$
& $0.7158 \pm 0.0013$ \\

Trop. + $L_{\mathrm{trop}}$
& $13.3670 \pm 0.0286$
& $0.9530 \pm 0.0020$
& $1.8681 \pm 0.0156$
& $\mathbf{0.7078 \pm 0.0039}$ \\

Trop. + $\ell_1 + L_{\mathrm{trop}}$
& $\mathbf{13.3330 \pm 0.0045}$
& $\mathbf{0.9506 \pm 0.0003}$
& $1.8300 \pm 0.0055$
& $0.7185 \pm 0.0042$ \\

Full model
& $13.3410 \pm 0.0241$
& $0.9511 \pm 0.0017$
& $\mathbf{1.8266 \pm 0.0041}$
& $0.7172 \pm 0.0023$ \\

Softmax + $L_{\mathrm{trop}}$ only
& $13.3450 \pm 0.0100$
& $0.9514 \pm 0.0007$
& $1.8352 \pm 0.0028$
& $0.7215 \pm 0.0016$ \\
\bottomrule
\end{tabular}

\vspace{0.8em}

\begin{tabular}{@{}lcccc@{}}
\toprule
Configuration
& MAE
& MRE
& Trop. dist.
& Ultra. viol. \\
\midrule
Trop. + $\ell_1$
& $\mathbf{0.0585 \pm 0.0040}$
& $\mathbf{0.1624 \pm 0.0077}$
& $0.2789 \pm 0.0087$
& $0.0507 \pm 0.0032$ \\

Trop. + $L_{\mathrm{trop}}$
& $2.0444 \pm 0.4196$
& $17.4117 \pm 3.6461$
& $0.3251 \pm 0.0077$
& $0.0520 \pm 0.0009$ \\

Trop. + $\ell_1 + L_{\mathrm{trop}}$
& $0.0597 \pm 0.0015$
& $0.2000 \pm 0.0074$
& $0.2597 \pm 0.0050$
& $0.0389 \pm 0.0049$ \\
Full model
& $0.0597 \pm 0.0018$
& $0.2163 \pm 0.0117$
& $\mathbf{0.2523 \pm 0.0086}$
& $\mathbf{0.0374 \pm 0.0026}$ \\

Softmax + $L_{\mathrm{trop}}$ only
& $0.1184 \pm 0.0137$
& $1.0240 \pm 0.1352$
& $0.2588 \pm 0.0054$
& $0.0390 \pm 0.0020$ \\
\bottomrule
\end{tabular}

\caption{
Ablation results on the held-out $10$-taxon test set.
Values are mean $\pm$ sample standard deviation across five training
seeds, after averaging over test alignments within each seed.
Lower values are better on all metrics. Bold indicates the best result in each column.
$L_{\mathrm{trop}}$ denotes the tropical symmetric loss.
}
\label{tab:ablation}
\end{table}

{The ablation study confirms that the $\ell_1$ term is necessary to minimize pairwise distance pair errors, while the tropical loss acts as a geometry-aware regularizer when combined with $\ell_1$. The full objective function with $\ell_1$, tropical, and ultrametric penalty provide the best overall balance across distance and branch-length-sensitive metrics, although the incremental effect of the ultrametric penalty is modest.}

\section{Conclusion}
We introduced a tropical axial attention model for phylogenetic tree inference on MSAs. The central motivation is that phylogenetic tree inference is not an unconstrained Euclidean regression problem. Instead, valid tree metrics on $n$ leaves occupy a structured, polyhedral subset of the ambient space $\mathbb{R}^{\binom{n}{2}}$. The tropical operations of the attention mechanism and the tropical metric in the loss function are designed to induce a bias in the model which is both accurate and consistent with tree space.

Across experiments with data generated under the same simulation regime as the training data, we see that the tropical model remains competitive while not outperforming the Phyloformer and Phyloformer 2 baseline models. When presented with data from a different simulation distribution, the tropical model improves, achieving topological results comparable to Phyloformer 2, outperforming both models on branch-length sensitive metrics, and providing a more consistent estimator for the tree metric induced by {FastME}.

The empirical experiments on DS1-DS11 reinforce this pattern. Because true empirical trees are unknown, these results do not directly measure the reconstruction accuracy. Instead, they show that the tropical model produces distance matrices that are much closer to their {FastME} reconstructed tree metrics than those produced by the comparison models. This suggests that the tropical axial attention model produces outputs more consistent with tree structure, even outside the simulated training distribution.

Overall, these results support the hypothesis that neural models benefit from aligning the inductive bias to the geometry of the problem. The piecewise linear operations of tropical axial attention provide an algebraic framework to learn the polyhedral structure of tree space. While the proposed model does not outperform across every metric, it shows robustness under distribution shift and improves tree metric consistency over the baseline models, indicating opportunities for further research in tropical attention models for structured inference problems.

{A practical limitation of the present implementation is its peak memory requirement during training. The tropical metric computation between the Q and K tensors in Equation \ref{eq:d_tr_qk} and the tropical matrix multiplication of these scores with the V tensor in Equation \ref{eq:attn_scores} are currently evaluated with Pytorch broadcast operations. This approach creates large intermediate tensors before applying the maximum and minimum reductions, substantially increasing GPU memory. More specialized implementations, such as the optimized tropical matrix-multiplication rountines in TropicalGEMM \citep{tropicalgemm}, may reduce this memory overhead if tensor broadcasts can be avoided.}

\bibliographystyle{plain}
\bibliography{refs}
\end{document}